# SOME MUSINGS ON GALAXY CLASSIFICATION


Sidney van den Bergh

Dominion Astrophysical Observatory

National Research Council

5071 West Saanich Road

Victoria, British Columbia

V8X 4M6

Canada






# ABSTRACT


The present paper presents a discussion of classification systems for galaxies, with special emphasis on possible modifications of the Hubble "tuning fork" diagram, and on galaxy types not included in Hubble's original scheme. Hubble's morphological types were defined in terms of standards observed at small look-back times that were mostly located in the field, or in poor clusters. It is pointed out that it is often difficult to shoehorn galaxies located in the cores of rich clusters, or objects viewed at large look-back times, into the Hubble classification system. The evolutionary relationships between E, S0 and dSph galaxies are presently still controversial and poorly understood. It is suggested that S0 galaxies may have arrived at their present morphology along various evolutionary tracks. Late-type barred spirals are found to be systematically less luminous than normal late-type galaxies. This suggests that the dichotomy between normal and barred spirals may reflect significant differences in their evolutionary histories. Such differences might be explored by searching for systematic differences between the [O/Fe] values in normal and barred spirals. Finally it is pointed out that the Large Magellanic Cloud may have been a low surface brightness galaxy for an ~8 Gyr period that ended 3-5 Gyr ago. This suggests that some galaxies can jump from one morphological classification type to another.




## 1.  INTRODUCTION

The present paper is mainly devoted to discussion of galaxies that are in the "ground state" (Ozernoy 1974).  Galaxies in "excited states" will not be mentioned in detail.  It is, however, noted that such excited galaxies occur in two families: (1) objects like Seyfert galaxies and quasars that are fueled by gas that is being fed into a compact <u>nucleus</u>, and (2) starburst galaxies in which gas is being fed into a galactic <u>disk</u>.  Current ideas on the classification of galaxies in the ground state are deeply rooted in concepts that were first developed by Edwin Hubble some eight decades ago.  In his pioneering paper Hubble (1926) arranged galaxies in his now famous "tuning fork" diagram

$$Sa \ - \ Sb \ - \ Sc$$
$$E0 - E7$$
$$SBa - SBb - SBc.$$

Later Hubble (1936) modified this scheme by incorporating a more-or-less hypothetical class S0 to span the chasm between elliptical and spiral galaxies, so that his modified tuning fork diagram took the form

$$Sa \ - \ Sb \ - \ Sc$$
$$E0 - E7 - S0$$
$$SBa - SBb - SBc.$$



Observations published by Sandage (1961) strengthened the justification for the introduction of type S0. Subsequently the Hubble/Sandage classification scheme was enhanced and expanded by Sandage (1975) and by Sandage & Tammann (1981). This work culminated in publication of <u>The Carnegie Atlas of Galaxies</u> (Sandage & Bedke 1994), which provides a magnificently illustrated exposition of the modified Hubble classification scheme.

A rather different approach to that adopted by Hubble and Sandage was chosen by Spitzer & Baade (1951), and by van den Bergh (1976), who viewed S0 galaxies as a sequence of flattened gas-free objects that parallels the Sa-Sb-Sc sequence for normal gas-rich galaxies. Finally Kormendy & Bender (1996) have recently suggested that the transition between ellipticals and early-type spirals may be provided by disk-like structures embedded in elliptical galaxies.

In recent years many authors have attempted to expand and embellish the original Hubble system. First Shapley & Paraskevopoulos (1940) sub-divided stage Sc into Sc and Sd, in which galaxies of stage Sd are later, i.e. have a patchier structure and more open spiral arms, than do objects of stage Sc. Subsequently de Vaucouleurs (1959a) introduced an additional classification stage Sm to mediate between types Sd and Im (in which the "m" denotes "magellanic", i.e. resembling



the Magellanic Clouds).  Furthermore Holmberg (1958) increased the resolution of the classification system for spirals by sub-dividing the main axis of the Hubble scheme into stages Sa-Sb$^-$-Sb$^+$-Sc$^-$ -Sc$^+$, with a parallel sequence for barred spirals. This work made it possible to establish a very close correspondence between the integrated colors of galaxies and their classification types.  This showed that there was a close correlation between galaxy classification type and the mean ages of the stellar populations that they contained.  An even more elaborate classification system was proposed by de Vaucouleurs (1959ab) who established a three-dimensional scheme with Hubble type E-Sa-Sb-Sc along the x coordinate, bar type SA-SAB-SB along the y axis, and arm shape r = (ring), rs = (mixed) and s = (spiral) along the z axis.

Shortly after Hubble (1926) published his galaxy classification system Reynolds (1927) commented that "[Hubble's] classification of spirals seems to me to be altogether too simple for the great range of types found... .  No classification would be complete unless development of the spiral form itself were taken into consideration".  An early attempt to carry out such a classification of spiral arms was made by Reynolds (1925) who noted that some galaxies had "massive" arms, whereas others exhibited more "filamentous" spiral arms.  Another attempt to classify galaxies on the basis of their arm morphology was made by Danver (1942).



More recently Elmegreen & Elmegreen (1982, 1987) have devised a twelve-stage classification system for spiral arms. These classifications range from Type 1 "flocculent" arms that are ragged, patchy, or chaotic to Type 12 "grand design" arms, which are long, symmetrical, sharply defined and dominate the appearance of the spiral galaxy in which they occur. A somewhat different approach was taken by van den Bergh (1960abc) who was able to show that the morphology of spiral arms correlates with the luminosity of their parent galaxy, in the sense that supergiant galaxies of luminosity class I have a pretty appearance that is dominated by long, well-defined spiral arms, whereas lower luminosity giants of luminosity class III have less well-defined patchy arms. The van den Bergh and Elmegreen classifications are loosely related in the sense that galaxies with patchy, fragmentary arms of Elmegreen Type 1 all have low luminosities, whereas grand design spirals of Elmegreen Type 12 are, without exception, objects of high luminosity.

A quite different approach to the classification of galaxy images was taken by Morgan (1958, 1959, 1962). Hubble had used both central concentration of light and the pitch angle of spiral structure, and its degree of resolution, to define his classification system. However, Morgan used only central concentration of light as a classification parameter. In the hands of Doi, Fukugita & Okamura



(1993) and Abraham et al. (1994) central concentration of light has been found to be a very useful parameter for computer-based classification of digital galaxy images.

## 2.      SOME PROBLEMS WITH THE HUBBLE SYSTEM

The discovery of dwarf spheroidal galaxies by Shapley (1939) showed that the original Hubble classification scheme did not encompass the entire realm of nebular morphologies.  It is now known that these faint dwarfs, which Hubble omitted from his classification system, are, in fact, the most numerous type of galaxy in the Universe.  At the other end of the galactic luminosity spectrum Hubble did not recognize bright cD galaxies (Matthews, Morgan & Schmidt 1964) as a separate morphological class.

### 2.1      The Hubble tuning fork diagram

The discoveries of cD and of dSph  galaxies serves to remind us of the fact that the Hubble (1936) galaxy classification system was defined in terms of a "training set" of bright giant and supergiant standards.  Strictly speaking the classical Hubble tuning fork diagram therefore applies only to luminous galaxies with $-22 < M_V < -18$.  Among subgiant galaxies with $M_V \sim -16$ the Hubble stages Sa, Sb and Sc begin to become meaningless.  For such objects it is only possible to



distinguish between early (smooth), intermediate, and late (patchy) spirals. Finally, among the least luminous objects near $M_V \sim -10$ it is only possible to recognize difference between dwarfs that are presently inactive and those that are among presently still forming stars.  It is now known that there is even an age hierarchy among the inactive dSph galaxies, with Ursa Minor and Draco systems containing mainly old stars, and Carina having a population that is mostly of intermediate age. Finally the Fornax dwarf spheroidal even contains some stars that are only a few x $10^8$ years old (Stetson 1997).  It is noted in passing that a similar age hierarchy exists among galaxies with $M_V \approx -16$.  The Andromeda companion M32 contains no young stars, NGC 205 exhibits some young blue stars, whereas the visual appearance of the Small Magellanic Cloud is dominated by such young stars.

Figure 1 presents a preliminary three-dimensional version of the Hubble

Place Figure 1 here

tuning fork.  This figure attempts to incorporate some of the effects of luminosity on galaxy morphology that have been discussed above.



2.2    Hubble stage versus luminosity

Table 1 and Figure 2 show a plot of the frequency distribution of

Place Table 1 and Figure 2 here

classification types  in A Revised Shapley-Ames Catalog of Bright Galaxies

(Sandage & Tammann (1991) as a function of absolute magnitude.  The figure

shows that (1) spiral galaxies along the sequence Sa-Sb-Sc have quite similar

luminosity distributions. However, the mean luminosity of late-type galaxies is seen

to drop precipitously along the sequence Sd-Sm-Im.  As a consequence the effects

of advancing classification stage, and of decreasing luminosity, are closely

intertwined in the classification systems of Hubble/Sandage and de Vaucouleurs.

2.3    The S0 class

The S0 class was introduced by Hubble (1936, p. 44) who wrote that

"nebulae intermediate between E7 and S0 are occasionally designated S0".  Baade

(1963, p. 78) stated that "In the end I think that it is quite clear that, if we

introduce the class S0, we should define it as the class of galaxies in which from

their general form we should expect to find spiral structure, but which, contrary to

our expectations, do not show it".  Sandage (1961) writes that the existence of S0

galaxies was established empirically by data accumulated during a study of

photographic survey of nearby galaxies carried out between 1936 and 1950.



However, Morgan (Matthews, Morgan & Schmidt 1964) does not appear to have been convinced. He writes that "the class S0, as used by Hubble, applies to galaxies having a variety of superficial appearances; that is, a mental picture of a unique galaxy form could not be derived from class S0". From a survey of available high quality surface photometry of early-type galaxies van den Bergh (1989) concluded that the photometric characteristics of these objects are only loosely correlated with their E or S0 classification by galaxy morphologists. Furthermore van den Bergh (1990) argues that the S0 classification type comprises a number of physically quite distinct types of objects that exhibit only superficial morphological similarities. Surveying all presently available data one has the impression that the S0 class represents a repository for objects that arrived at their present morphology via quite different evolutionary paths. For example, some of the S0 galaxies in rich clusters might have started off as spirals with big bulges (à la NGC 4594 = M104 "the Sombrero") that were swept free of gas by ram-pressure stripping (Gunn & Gott 1972), whereas the gas and dust in others, like the field S0 NGC 3115, might have been blown out when the $1 \times 10^9$ $M_\odot$ black hole in its nucleus (Kormendy & Richstone 1992) was a quasar. Other S0/SB0 galaxies, such as NGC 5495 (the companion to "the whirlpool nebula" M51) and NGC 5128 (= Cen A), may owe their present morphologies to star formation induced by relatively recent tidal captures of significant amounts of gas.



A strong argument against the hypothesis that S0 galaxies form a bridge that spans the chasm between E and S0 galaxies is provided by the observation (van den Bergh 1990) that S0 galaxies are, in the mean, significantly fainter than both ellipticals and spirals of type Sa. Alternatively one might make the <u>ad hoc</u> assumption that there are two distinct classes of S0 galaxies; one of which is truly intermediate between types E and Sa, whereas the other consists of lenticular objects that are much fainter than both ellipticals and early-type spirals. Possibly some S0 galaxies with a bright spheroid and a faint disk have been mis-classified as ellipticals.

Sandage (1961) has sub-divided S0 galaxies into sub-types $S0_1$, $S0_2$ and $S0_3$, on the basis of the presence (or absence) of dust lanes. The different sub-types of the S0 class are found to have luminosity distributions that are indistinguishable from each other. Furthermore no dependence of (projected) galaxy flattening is found among members of the S0 class. The latter result is unexpected because the luminosity distribution of ellipticals peaks ~1.5 mag brighter than that of S0 galaxies (van den Bergh 1990). One might, perhaps, have expected round S0's to have had luminosities similar to those of ellipticals. Sandage (1961) speculated that those Sa galaxies that have spiral arms [Sa(<u>s</u>)] constitute a continuation of type $S0_2$, whereas he hypothesizes that Sa galaxies



with rings [Sa($\underline{r}$)] branch off from type S0$_3$.  An argument against this hypothesis

is, however, provided by the finding (van den Bergh 1998) that S0$_2$ galaxies are, in

the mean, significantly fainter than Sa($\underline{s}$) galaxies.  Furthermore the luminosity

distribution of Sa($\underline{r}$) galaxies peaks at a higher luminosity than does that of S0$_3$

galaxies.

2.4     How to tell late-type spirals from irregulars

        The Hubble classification scheme does not provide an entirely objective

criterion for distinguishing irregular galaxies from late-type spirals.  A particular

problem is provided by NGC 4449.  This object was used by Hubble (1936) as the

prototype for the class of irregular galaxies.  Sandage (1961) concurred with this

classification.  However, Sandage & Tammann (1981) assign NGC 4449, and

many other galaxies such as the Large Magellanic Cloud (LMC), that had

previously been regarded as irregulars, to class Sm.  Perhaps the most objective

way to avoid this difficulty is (Hubble, 1936, p. 47, van den Bergh 1995) to use the

presence (or absence) of a nucleus as the criterion for distinguishing spiral galaxies

from irregulars.  Using this criterion the Large Magellanic Cloud is of type IBm,

rather than of type SBm.  It is noted parenthetically that Hubble (1936) ignored the

fact that the dichotomy between normal and barred spirals continues into the

domain of the irregular galaxies, with the SMC being an example of a normal



irregular whereas the LMC is a prototypical barred irregular.

Holmberg (1958) first made a distinction between galaxies of Type Ir I, which are now mostly referred to as Im, and objects of Type Ir II, many of which appear to have been galaxies of types E-Sa-Sb that recently captured large amounts of gas and dust. Alternatively some Ir II galaxies may be spirals that have undergone violent tidal interactions in the recent past. NGC 3077 is an example of an early-type ancestor that was probably transformed into an Ir II galaxy by capture of gas and dust. On the other hand NGC 3034 (= M82) may have been a late-type galaxy that captured gas from NGC 3031 (= M81). Finally NGC 520 is the prototype of an Ir II galaxy that almost certainly resulted from a violent tidal interaction.

2.5     Rich clusters and large look-back times

The Hubble scheme was developed to provide a convenient way to describe the morphological types of nearby galaxies. It is therefore not surprising that it provides us with a powerful tool for the classification of galaxies (with bright apparent magnitudes) which are predominantly situated in the field or in small clusters. However, the Hubble system proves to be a rather blunt tool for the classification of galaxies in the cores of rich clusters. The majority of such



objects appear to be of types E, S0, and SB0. For such galaxies a simple classification scheme à la Morgan, based on the central concentration of light in galaxy images, appears to provide a satisfactory classification (Abraham et al. 1994). A similar dilemma is encountered in the Hubble Deep Field (Abraham et al. 1996, van den Bergh et al. 1996) in which it turns out to be difficult to "shoehorn" a large fraction of the galaxy images into the morphological categories provided by the Hubble scheme. Presumably this difficulty is mostly due to the fact that (1) many disk galaxies have not yet had time to assemble from ancestral objects, (2) interactions are more frequent than they are among galaxies at lower redshifts, and (3) for reasons that are not yet understood few, if any, very young spirals in the Hubble Deep Field appear to exhibit bars.

## 3. MORGAN'S CLASSIFICATION SYSTEM

The galaxy classification system developed by Morgan (1958, 1959, 1962), which is sometimes referred to as the Yerkes system, is a one-dimensional scheme based entirely on the central concentration of light in a galaxy image. Galaxies are arranged in the sequence a-af-f-fg-g-gk-k in which objects of type a have the lowest central concentration and those of type k have the highest central concentration of light. The corresponding integrated galactic spectral types range from A to K. A classification system based on central concentration of light in



galaxy images is particularly suitable for automatic computer-based study of galaxies (Abraham et al. 1994). In addition to central concentration, Morgan also assigned galaxies to the form families E (elliptical), S (spiral), B (barred), I (irregular), L (low surface brightness), D (dustless) and N (objects with a bright nucleus superimposed on a fainter background). The Hubble/Sandage class S0 corresponds approximately to de Vaucouleurs' lenticular type and to Morgan's D form family. A problem with Morgan's D type is, however, that the distinction between his E and D families appears somewhat artificial and ill-defined.

A comparison between galaxy classifications on the Hubble (Sandage & Tammann 1981) and Morgan (1959) systems is given in Table 2 and is shown in

Place Table 2 here

Fig. 3. This figure shows that the Yerkes system has relatively low resolution for

Place Figure 3 here

Hubble types E, S0 and Sa, which are mostly of concentration class $\underline{k}$. On the other hand galaxies of Hubble stage Sc are seen to exhibit a large dispersion in their Morgan concentration type indices. Perhaps surprisingly, no systematic luminosity differences are found between the diffuse Sc galaxies of types $\underline{a}$ and $\underline{af}$, and the more compact ones of types $\underline{fg}$ and $\underline{g}$. Finally there appears to be no systematic difference between the relationship of the Hubble and Morgan systems



for field galaxies on the one hand, and for galaxies in the Virgo cluster on the other.

## 4.    GALAXY EVOLUTION AND CLASSIFICATION

Morgan (1958) has emphasized that "The value of a classification system depends on its usefulness". The usefulness of the morphological classifications discussed above is most clearly demonstrated by the strong correlation between integrated colors and classification types (Holmberg 1958, de Vaucouleurs 1959b, 1963). The existence of these correlations shows that a deep linkage must exist between morphology and stellar population content along the classification sequence E-Sa-Sb-Sc-Sd-Sm-Im. On the other hand de Vaucouleurs (1961) finds no significant differences between the integrated colors of normal and barred spirals of the same Hubble stage. This might lead one to suspect that the dichotomy between normal and barred spirals does not correlate with profound differences in their stellar populations, and hence in their evolutionary histories. However, Figure 4 and the data in Table 3 show that the luminosity distribution of

Place Figure 4 and Table 3 here

normal galaxies of type Sc differs significantly from that of barred galaxies of stage SBc, in that barred objects are systematically fainter than normal disk galaxies having the same Hubble stage. A Kolmogorov-Smirnov test shows that there is



only a ~0.2 percent probability that the Sc and SBc galaxies in Sandage &

Tammann (1981) were drawn from the same parent luminosity distribution. In van

den Bergh (1998) it is shown that this difference is unlikely to be due to small

systematic differences in the stage assignment of normal and barred spirals. If this

conclusion is correct then there must be some profound difference between the

evolutionary history of normal and barred late-type spirals, which results in barred

objects having systematically lower luminosities than normal ones. How might this

difference be accounted for? Noguchi (1996) has recently suggested that slow

infall from a protogalactic halo will keep the gas mass in a galactic disk low,

resulting in a low star formation rate, and hence in a cold disk that can develop a

bar-like instability. On the other hand rapid infall from a massive gaseous halo

might result in a pile-up of massive gas clumps in the disk, producing rapid star

formation which would make the disk dynamically hot. The resulting large random

motions might suppress bar-like disk instabilities. If this scenario is correct then

barred late-type spirals may have collapsed more slowly than normal late-type

galaxies. On this hypothesis the [O/Fe] ratio would be expected to be higher in

normal spirals than it is in barred spirals. This is so because short-lived supernovae

of Type II would contribute more to the enrichment of the interstellar medium in

fast-collapsing normal spirals, than they would in more slowly evolving proto

barred spirals. On the other hand enrichment by the more slowly evolving



supernovae of Type Ia would be more important for enrichment of barred spirals.

Unfortunately available observations (Jablonka, Martin & Arimoto 1996) are

presently not yet suitable for performing this test.  A possible complication is that

the presence of a bar may stir up disk gas, and therefore erase pre-existing O and

Fe gradients.  On the other hand the existence of such a bar might also feed gas

into the center of a galaxy, possibly resulting in a burst of nuclear star (and SNe II)

formation, which might enhance pre-existing stellar age and metallicity gradients.

## 5.        LOW SURFACE BRIGHTNESS GALAXIES

Galaxies on the Hubble sequence Sa-Sb-Sc have central disk surface

brightnesses that appear to fall in a rather narrow range (Freeman 1970).  Disney

(1976) has emphasized the fact that the scarcity of low surface brightness disk

galaxies in most catalogs is due to selection effects, which discriminate against low

surface brightness (LSB) galaxies.  It now appears that such LSB galaxies can be

grouped into four broad classes:  (1) Monsters, such as Malin #1 (Bothun et al.

1987, Impey & Bothun 1989), with dimensions that are comparable to those of the

cores of clusters of galaxies.  Such objects might be the remnants of cosmic

catastrophes that took place in the distant past.  (2) LSB galaxies with luminosities

similar to those of most normal galaxies.  The fact that such dwarf galaxies are

generally quite blue rules out the possibility (McGaugh 1992) that LSB galaxies



are normal disk galaxies that have faded as their stars aged and evolved. (3)

Dwarf irregular (dIm) and dwarf spheroidal (dSph) galaxies. Such dSph and dIr

galaxies probably differ only in their star formation history. (4) Low surface

brightness galaxies in rich clusters. Such objects, of which NGC 4411A and NGC

4411B in the Virgo cluster are the prototypes (van den Bergh 1998), probably

have a low rate of star formation because their gaseous disks were depleted by

ram-pressure stripping (Gott & Gunn 1972).

Observation of the Large Magellanic Cloud suggest that LSB galaxies can

transform themselves into normal disk galaxies. From observations of LMC field

stars near the main sequence turnoff point of Population II Butcher (1977) was

able to show that a major burst of star formation, which extends to the present

day, started in the Large Magellanic Cloud 3-5 Gyr ago. This unexpected

conclusion was subsequently confirmed by Stryker (1983) and by Hardy et al.

(1984). From an investigation of three starfields in the Large Magellanic Cloud

Bertelli et al. (1992) estimate that the mean rate of star formation in the LMC was

as much as ten times lower before the burst, than it has been since then. A rather

different conclusion has recently been obtained by Holtzman et al. (1997) who,

from the study of a single field in the outer region of the Large Cloud, conclude

that the best fit to the observed stellar luminosity function is provided by a model



in which the rate of star formation in the LMC was, during most of its lifetime, only about three times lower than it has been since the great burst of star formation that started a few Gyr ago. Caveats are that such calculations depend in a rather sensitive way on the adopted luminosity function of star formation and, to a lesser extent, on details of the assumed heavy element enrichment history of the Large Cloud. Da Costa (1991) has shown that a similar hiatus exists between the ages of the 13 LMC globular clusters, which formed ~13-15 Gyr ago, and Large Cloud open clusters that started to form 3-5 Gyr ago. These observations suggest that the LMC experienced a "dark age" that lasted ~8 Gyr during which the rate of star and cluster formation may have been as much as an order of magnitude lower than it is at present. During these dark ages the Large Cloud would have had the characteristics that are associated with typical LSB galaxies. It is therefore concluded that the LMC provides prima facie evidence in favor of the hypothesis that galaxies can jump from one morphological type to another. The reason for the transition of the Large Cloud from classification type LSB to IBm remains a mystery. The fact that The Small Magellanic Cloud does not exhibit a peak in its cluster formation rate 3-5 Gyr ago (Da Costa 1991) appears to rule out the suggestion that the LMC starburst was triggered by a tidal encounter between the Large and Small Magellanic Clouds.



The Large Cloud presently has a low specific globular cluster frequency of S = 0.5: (Harris 1991). During the "Dark Ages" the low surface brightness of the LMC would have given the Large Cloud a much higher value of S. This suggests that it might be of interest to see if relatively nearby LSB galaxies also have an unexpectedly high specific globular cluster frequency.

## 6. CONCLUSIONS

Present thinking about the morphological classification of galaxies remains firmly rooted in ideas introduced by Hubble (1926, 1936). It has, however, become clear that the Hubble scheme is only strictly applicable to galaxies with absolute magnitudes in the range $-22 < M_V < -18$ that are observed at small look-back times in the field or in small clusters. It is not possible to shoehorn a large fraction of the galaxies seen at large look-back times, and in the cores of rich clusters, into the morphological types defined by Hubble. Additional problems are that galaxies of type S0 appear to be a mixture of objects that have arrived at their present morphology via a number of quite different evolutionary paths. Both the notion that S0 galaxies form a bridge that spans the chasm between ellipticals and spirals, and the suggestion that S0 galaxies form a sequence that parallels that of normal spirals, therefore probably represent an over-simplification. It is pointed out that there are serious inconsistencies between the assignments of galaxies to



the Sm and Im classes. This problem may be avoided by using the presence (or absence) of a nucleus as the criterion for distinguishing between spirals and irregulars. The fact that late-type barred spirals are systematically fainter than normal spirals of similar Hubble stage suggests that there might be significant differences between the evolutionary histories of normal and barred spirals. It would be of interest to search for such systematics by looking for differences between the oxygen-to-iron ratios in normal and barred spirals. Finally it is suggested that the classification type of Large Magellanic Cloud may have jumped from LSB to IBm 3-5 Gyr ago.

**TABLE 1**

**Normalized frequency distribution of spiral galaxies in the Revised Shapley-Ames Catalog (Sandage & Tammann 1981)[a]**

| $M_B$ | Sa | Sab | Sb | Sbc | Sc | Scd | Sdm | Sm+Im |
|---|---|---|---|---|---|---|---|---|
| -23.75 | ... | ... | 0.01 | ... | ... | ... | ... | ... |
| -23.25 | 0.01 | 0.03 | 0.02 | 0.04 | 0.01 | ... | ... | ... |
| -22.75 | 0.03 | 0.10 | 0.19 | 0.01 | 0.03 | ... | ... | ... |
| -22.25 | 0.13 | 0.10 | 0.12 | 0.12 | 0.05 | ... | ... | ... |
| -21.75 | 0.13 | 0.15 | 0.24 | 0.22 | 0.17 | ... | ... | ... |
| -21.25 | 0.29 | 0.21 | 0.13 | 0.29 | 0.16 | ... | ... | ... |
| -20.75 | 0.17 | 0.21 | 0.14 | 0.14 | 0.20 | 0.08 | ... | ... |
| -20.25 | 0.16 | 0.21 | 0.07 | 0.10 | 0.15 | 0.15 | ... | 0.10 |
| -19.75 | 0.08 | ... | 0.04 | 0.04 | 0.12 | 0.23 | 0.08 | 0.10 |
| -19.25 | 0.03 | ... | 0.02 | 0.04 | 0.06 | 0.15 | 0.17 | ... |
| -18.75 | ... | ... | 0.01 | 0.01 | 0.03 | 0.23 | 0.17 | 0.10 |
| -18.25 | ... | ... | ... | ... | 0.01 | 0.08 | 0.17 | 0.10 |
| -17.75 | ... | ... | ... | ... | ... | 0.08 | 0.17 | ... |
| -17.25 | ... | ... | ... | ... | ... | ... | 0.08 | 0.20 |
| -16.75 | ... | ... | ... | ... | ... | ... | ... | 0.10 |
| -16.25 | ... | ... | ... | ... | ... | ... | 0.08 | 0.10 |
| -15.75 | ... | ... | ... | ... | ... | ... | ... | ... |
| -15.25 | ... | ... | ... | ... | ... | ... | ... | 0.20 |
| -14.75 | ... | ... | ... | ... | ... | ... | 0.08 | ... |

[a]    Due to rounding errors not all columns add up to 1.00

# TABLE 2

## Comparison between Morgan and Hubble classifications

|      | E  | E/S0 | S0 | S0/Sa | Sa | Sab | Sb | Sbc | Sc | Scd | Sd | Sm | Ir |
|------|----|------|----|-------|----|-----|----|-----|----|-----|----|----|----|
| *k*  | 69 | 12   | 53 | 15    | 17 | 6   | 4  | 0   | 0  | 0   | 0  | 0  | 0  |
| *gk* | 2  | 0    | 8  | 1     | 8  | 3   | 11 | 1   | 0  | 0   | 0  | 0  | 0  |
| *g*  | 0  | 0    | 4  | 0     | 9  | 8   | 24 | 9   | 3  | 1   | 0  | 0  | 0  |
| *fg* | 0  | 0    | 0  | 0     | 3  | 1   | 5  | 13  | 11 | 0   | 0  | 0  | 0  |
| *f*  | 0  | 0    | 0  | 0     | 1  | 1   | 11 | 15  | 41 | 1   | 1  | 0  | 0  |
| *af* | 0  | 0    | 0  | 0     | 0  | 0   | 0  | 9   | 40 | 2   | 0  | 0  | 0  |
| *a*  | 0  | 0    | 0  | 0     | 0  | 1   | 0  | 2   | 29 | 2   | 2  | 3  | 1  |

**TABLE 3**

**Luminosity distributions for Sc and SBc galaxies in the Shapley-Ames catalog.**[a]

| $M_B$ | $N(M_B)^b$ | $N(M_B)^c$ |
|-------|-----------|-----------|
| -23.25 | 4 | 0 |
| -22.75 | 8 | 1 |
| -22.25 | 15 | 3 |
| -21.75 | 48 | 4 |
| -21.25 | 45 | 12 |
| -20.75 | 58 | 9 |
| -20.25 | 43 | 10 |
| -19.75 | 34 | 21 |
| -19.25 | 18 | 8 |
| -18.75 | 8 | 3 |
| -18.25 | 4 | 2 |

[a]   from Sandage & Tammann (1981)

[b]   ordinary spirals

[c]   barred spirals

**FIGURE CAPTIONS**

Fig. 1   Tentative proposal for a three-dimensional "tuning fork" diagram.

Fig. 2   Luminosity distribution for Shapley-Ames galaxies (Sandage & Tammann 1981) as a function of Hubble type.  Note that objects of types Sd-Sm-Im are, on average, significantly less luminous than spirals of type Sa-Sb-Sc.

Fig. 3   Comparison between the Hubble/Sandage and Morgan classifications.  The figure shows that the Yerkes system has low resolution for Hubble types E-S0-Sa, and that galaxies of stage Sc have a wide range in central concentration of light.

Fig. 4   Comparison of the luminosity distributions of normal and barred galaxies of stages Sc-Sd-Sm-Im in the Shapley-Ames Catalog (Sandage & Tammann 1981).

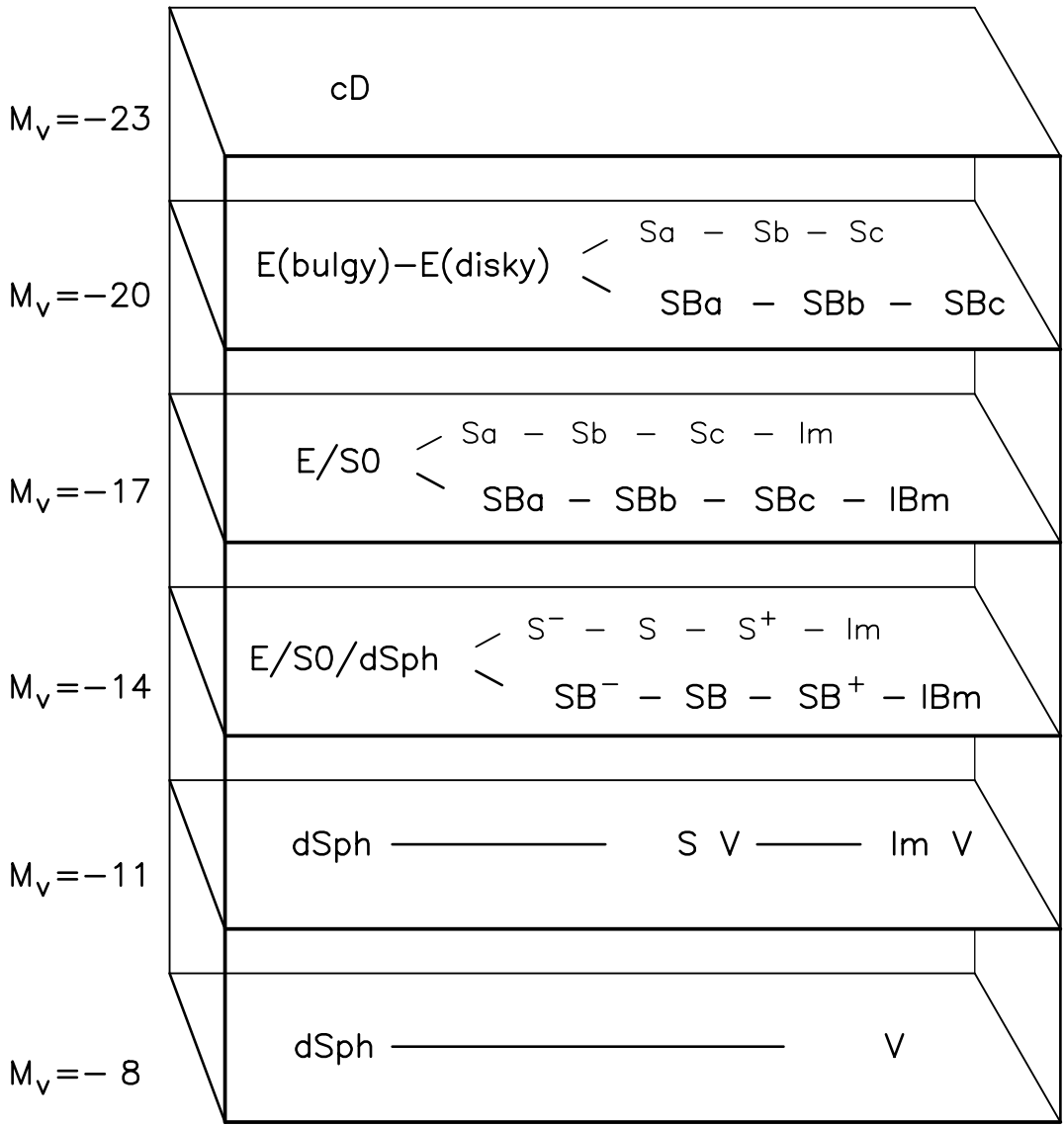

$M_v = -23$

cD

$M_v = -20$

E(bulgy)−E(disky)  Sa − Sb − Sc
                   SBa − SBb − SBc

$M_v = -17$

E/S0  Sa − Sb − Sc − Im
      SBa − SBb − SBc − IBm

$M_v = -14$

E/S0/dSph  S⁻ − S − S⁺ − Im
           SB⁻ − SB − SB⁺ − IBm

$M_v = -11$

dSph ——————— S V ——— Im V

$M_v = -8$

dSph ——————— V

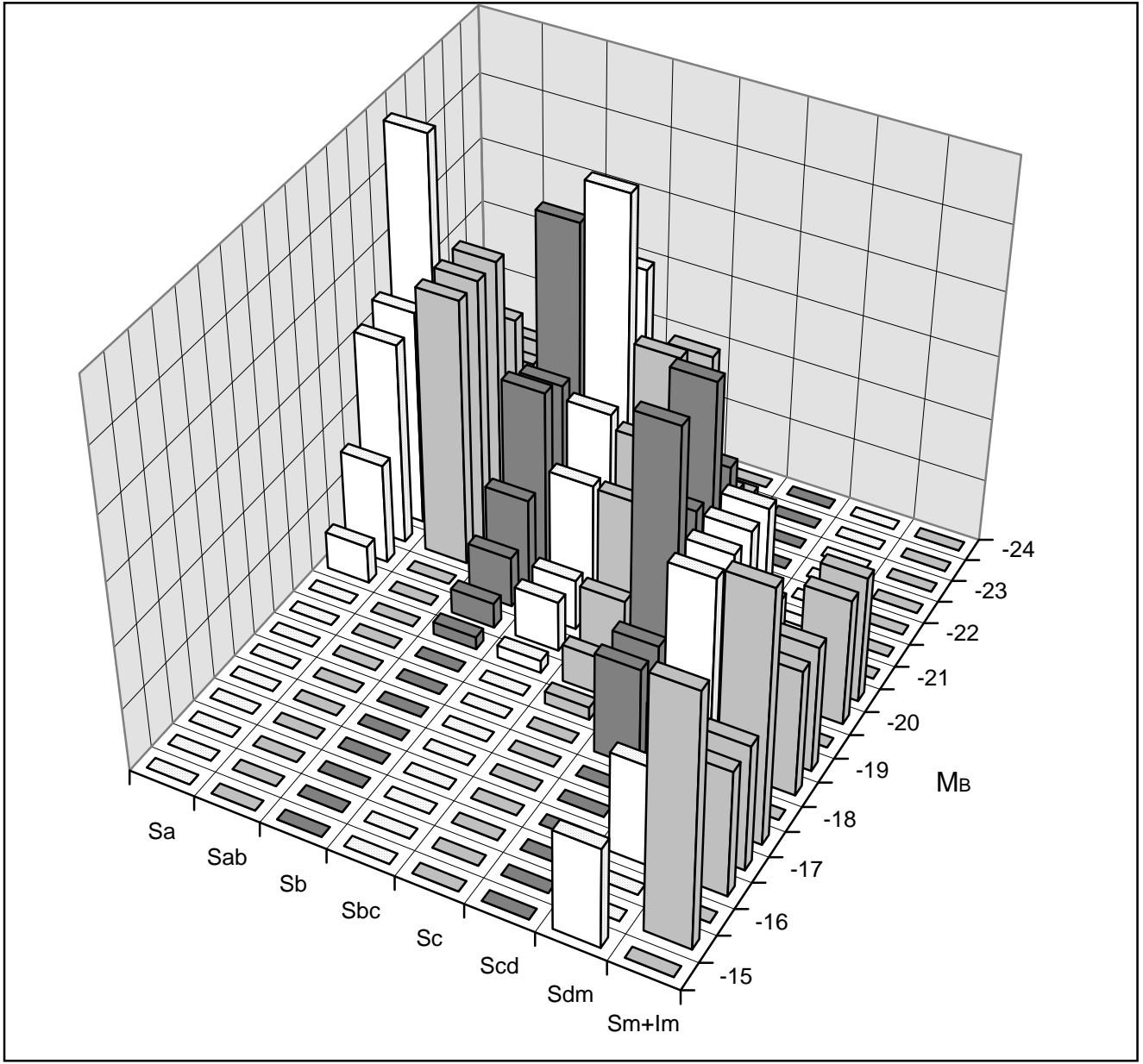

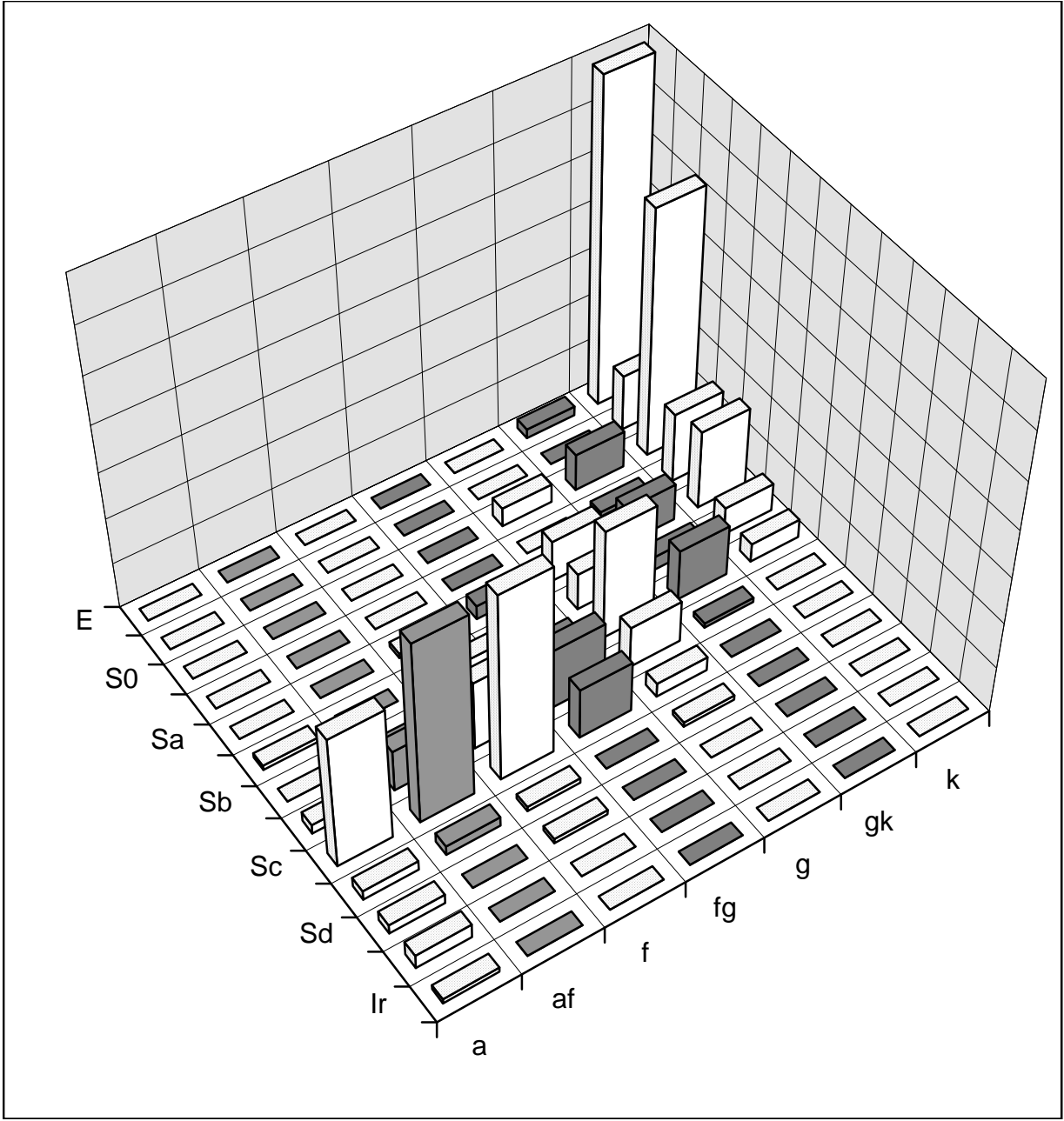

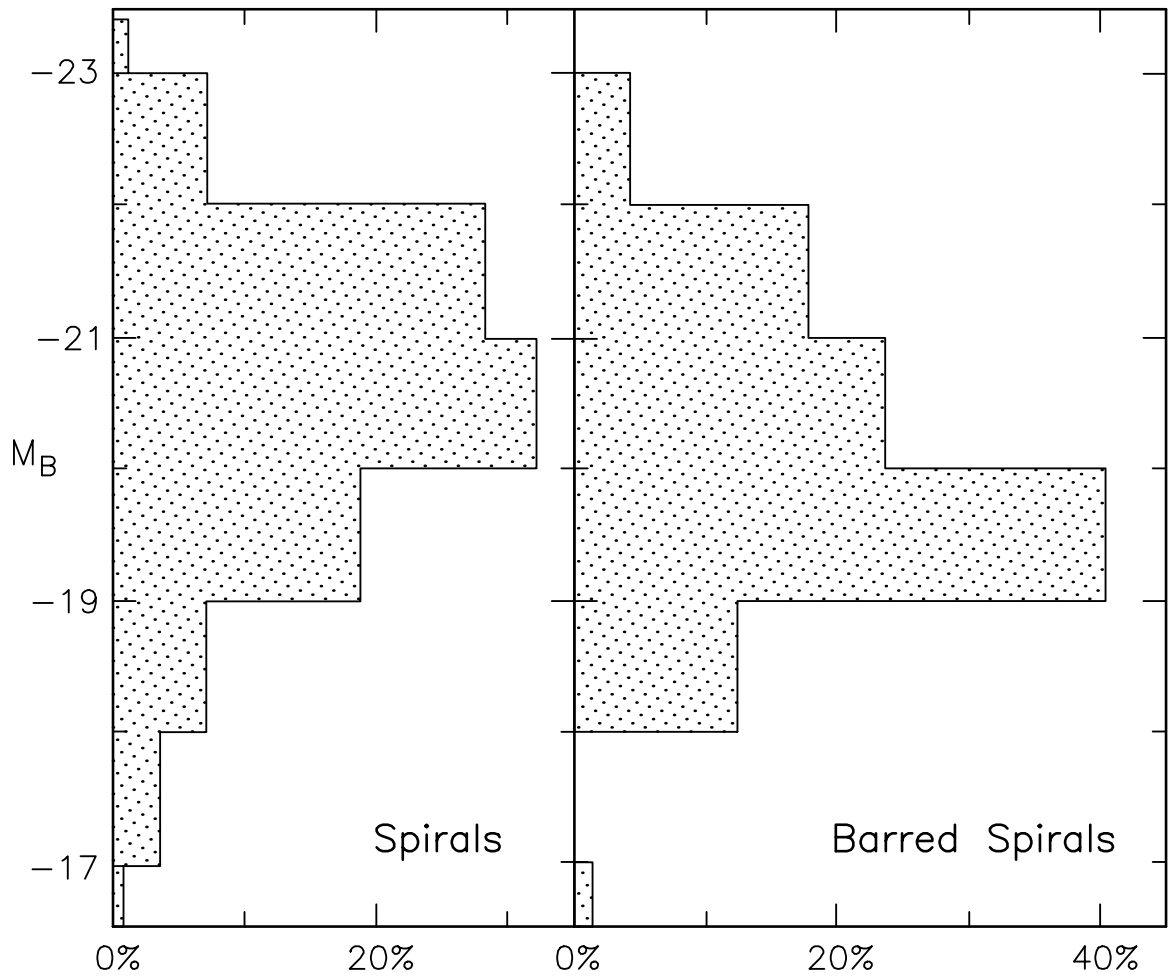